\documentclass[%
 reprint,
 amsmath,amssymb,
 aps,
]{revtex4-2}

\usepackage{graphicx}
\usepackage{dcolumn}
\usepackage{bm}

\begin{document}

\preprint{APS/123-QED}

\title{Chaos from a free-running broad-area VCSEL}
\author{Jules Mercadier$^{*,1,2}$}

\author{Stefan Bittner$^{1,2}$}
\author{Damien Rontani$^{1,2}$}%
\author{Marc Sciamanna$^{1,2}$}
\affiliation{%
 $^1$Université de Lorraine, CentraleSupélec, LMOPS EA-4423, 57070 Metz, France\\
 $^2$Chaire Photonique, LMOPS EA-4423, CentraleSupélec, 57070 Metz, France \\
 $^*$ Corresponding author : jules.mercadier@centralesupelec.fr
}%

\begin{abstract}
We experimentally report on the detection of chaos from a free-running commercial broad-area VCSEL without the need for external perturbation such as optical feedback, injection or current modulation. The evolution of nonlinear dynamics leading to chaotic behavior is studied, and the system's complexity is characterized using chaos titration and correlation dimension. We link the occurence of chaos with the complex interplay between the spatial laser modes competition and polarization dynamics.
\end{abstract}

\maketitle

\section{Introduction}

Understanding the nonlinear dynamics of semiconductor lasers is key to advancing all-optical information processing technologies, particularly with the promising applications of optical chaos \cite{sciamanna_physics_2015}. Current methods for generating optical chaos typically involve external perturbations, such as feedback, optical injection, or parameter modulation \cite{Kovanis_1995_Instabilitites, lang_external_1980}. However, these approaches often face challenges like time-delay signatures (TDS) \cite{rontani_loss_2007}, complex setups, and stability issues.

Among semiconducor lasers, vertical-cavity surface-emitting lasers (VCSELs) exhibit significant potential for applications due to their low threshold, surface-emitting geometry, which allows for a high density of lasers on a small surface area, as well as their low power consumption. Due to their particular geometry, VCSELs can exhibit not only linear but also circular or elliptic polarization states with the possibility of polarization switching when varying the laser parameters. Polarization competition is often accompanied by complex nonlinear dynamics \cite{olejniczak_polarization_2011,van_exter_polarization_1998}. The polarization properties of each of the individual VCSEL lasing modes has also been studied  \cite{review_polar_dynamic, review_wang_APL} and spatio-temporal dynamics is predicted to emerge due to the presence of multiple transverse modes \cite{martin-regalado_polarization_1997,valle_dynamics_1995}. All of these characteristics suggest that VCSELs could function as standalone chaos generators, without the need for external perturbations.

Several studies, for instance, have demonstrated polarization chaos generation in VCSELs \cite{virte_deterministic_2013, virte_bifurcation_2013}. While extensive research has been conducted on single-mode VCSELs, broad-area VCSELs remain less well understood \cite{buccafusca_transient_1999,richie_chaotic_1994,valle_dynamics_1995,GIUDICI1998313}. Previous studies have not yet provided a comprehensive understanding or sufficient evidence for the existence of chaos in broad area VCSEL. Recent findings suggest complex spatio-temporal dynamics \cite{bittner_complex_2022}, hence motivating further experimental investigations of these lasers.

In this paper, we perform an in-depth analysis of the dynamic of a commercial broad-area VCSEL (BA-VCSEL). We highlight the generation of complex nonlinear dynamics, some of which are identified as deterministic chaos. Varying the injection current reveals a sequence of self-sustained periodic and chaotic dynamics, along with polarization switching and spatial mode redistribution. Moreover, our analysis demonstrates a higher correlation dimension of the chaotic dynamics compared to the case of chaos from a single-mode VCSEL \cite{virte_deterministic_2013}.
\section{Experimental Setup}
\begin{figure}
    \centering
    \includegraphics[scale=0.6]{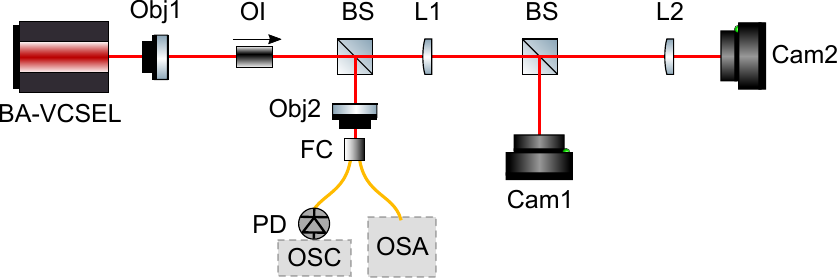}
    \caption{Experimental setup. Obj1 : 40x objective (NA = 0.6); OI : optical isolator; BS : beam splitter; FC : fiber coupler; OSC : oscilloscope; OSA : optical spectrum analyzer; L1 (L2) : lens with 200mm (100mm) focal length; Cam1 (Cam2) : CCD Camera for near-field (far-field) image of VCSEL; Obj2 : 20x objective (NA = 0.5).}
    \label{setup experimental}
\end{figure}

\begin{figure*}
    \includegraphics{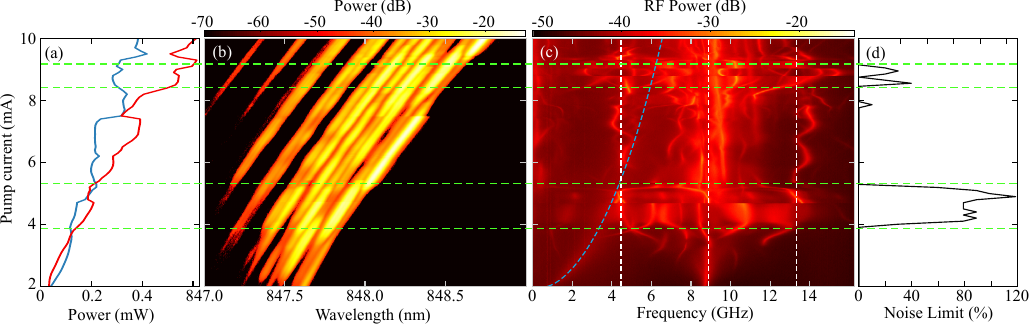}
    \caption{(a) LI-curves for the two main polarization axes $\alpha_u$ = 0° (blue) and $\alpha_v$ = 90° (red). (b) Evolution of the laser spectrum as function of the pump current for $u$-polarization. (c) RF-spectra as function of the pump current. The vertical white dashed line indicate the birefringence frequency $f_b$ as well as 0.5$f_b$ and 1.5$f_b$. The blue curved dashed line indicates the relaxation oscillation frequency $f_{RO}$. (b) inset : spatio-spectral image of the BA-VCSEL at 4.9 mA. (d) Chaos titration. The green dashed lines delimit the region with a noise limit $>$ 0\% indicating chaotic dynamic.}
    \label{fig_principal}
\end{figure*}
Figure \ref{setup experimental} presents a schematic representation of our experimental setup. The laser under study is a broad-area VCSEL with a circular output aperture of 15 $\mu m$ in diameter (Frankfurt Laser Company FL85-F1P1N-AC) pumped through a ring contact. The laser mount stabilizes the temperature at 20°C using a thermo-electric cooler. The laser output is collimated using a 40x microscope objective (Obj1). An optical isolator (OI) with an adjustable mount is used to select the polarization angle and to prevent any back reflections into the VCSEL. The orientation of the isolator also plays a crucial role in selecting the polarization at the laser output. In our case, with both polarizations being linear and orthogonal, we can effectively retrieve part of the signal for both polarization orientations (0° and 90°) by aligning the isolator at 45°. Using a beam splitter (BS), part of the signal is directed to a fiber-coupler (FC) with a 20x objective (Obj2) to couple it into a multimode fiber (Thorlabs M116L02), which is connected either to a photodetector (Newport 1484-A-50, 22 GHz) connected to an oscilloscope (Tektronix DPO72340SX - 23 GHz bandwidth) or to an optical spectrum analyzer (Anritsu MS9740A). The objective lens has a sufficient aperture and angular acceptance to collect the entire signal and match the fiber’s numerical aperture, ensuring that all spatial modes of the fiber are properly captured. The signal from the photodetector is amplified using an RF-amplifier (SHF S126A, 29 dB). The other part of the laser emission is sent to a 200 mm focal length lens (L1) to capture near-field images on a CCD camera (Cam1). A second CCD camera (Cam2) records the far-field images using a lens with a 100 mm focal length (L2). This setup allows us to monitor the temporal dynamics ans the optical spectra of the laser, and we perform automated measurements to reveal dynamical transitions as the laser pump current is varied. \\
\section{Results}

First, we investigate the polarization state of the VCSEL. The transmission of the laser through a rotating polarizer with angle $\alpha$ varying from 0° to 180° in 10° increments was measured between 0 and 10 mA. The results are shown in Fig. \ref{fig_principal}(a): The laser exhibits a threshold current of $I_{th}$=1.8 mA and two orthogonal main polarization axes, $u$ and $v$ with $\alpha_u$ = 0° (blue curve) and $\alpha_v$ = 90° (red curve). As the pump current increases, the dominant polarization axis changes several times, a phenomenon known as polarization switching which is well documented for single-mode VCSELs \cite{martin-regalado_polarization_1997, virte_deterministic_2013}. \\
These polarization switching points (PSPs) result from mode competition \cite{bittner_complex_2022}. Figure \ref{fig_principal}(b) shows the evolution of the spectrum as a function of the pump current, with the angle of the polarizer of the optical isolator fixed along the $u$-axis. As the pump current increases, we observe a red-shift of all modes due to Joule heating. Additionally, we note that the distribution of power among the modes changes between the different transverse modes and the two polarization states, which leads to a PSP when the total power of modes in one polarization overtakes that in the other polarization. The inset in \ref{fig_principal}(b) shows the spatio-spectral image of the BA-VCSEL at 4.9 mA for the $u$-axis. It is measured using a spectrometer (Princeton Instruments SpectraPro HRS-500) with a 1800 g/mm grating. We observe the presence of three main modes which is also visible in the optical spectra along with a few additional weaker modes which are partially or not visible on the spectrometer. The three main visible modes can be identified as the modes (1,1), (3,1), (4,1).\\
The dynamics of this system has been thoroughly studied in Ref.\cite{bittner_complex_2022}, highlighting the challenges of developing a theoretical model for this type of VCSEL. It is understood that the complex multimode interactions, combined with the polarization dynamics, lead to changes in the temporal dynamics, as visible in the frequency domain in Fig. \ref{fig_principal}(c). One of the system's characteristic frequencies that can be identified is the birefringence. The optical spectrum analysis allows to determine the laser birefringence by measuring the frequency shift between the orthogonal polarization modes, here approximately 9 GHz. Since birefringence cannot be measured using the resolution of our OSA, this measurement was carried out by separately taking two optical spectra at the same current, for the two polarization axes $u$ and $v$, and comparing them to determine the birefringence value. The RF spectrum contains several frequency components. As the pump current increases, additional frequency components emerge or vanish, along with regions in which the frequencies of certain components drift. Most frequency components are found around the birefringence frequency $f_b$, its subharmonics 0.5$f_b$ as well as 1.5$f_b$ (indicated by the white dashed lines). The creation of additional frequency components as the pump current increases suggests transitions between different dynamic regimes, and potentialy the emergence of chaotic dynamics. The transitions in the dynamics coincide with significant changes in the optical spectrum [Fig. \ref{fig_principal}(b)] and PSPs [Fig. \ref{fig_principal}(a)], which highlights how the mode competition impacts the dynamics and stability of the system. In the following we will analyse the different dynamic regimes using the well-established tools of time series analysis, previously applied to single-mode VCSELs \cite{virte_deterministic_2013}.

\begin{figure}[!ht]
    \centering
    \includegraphics{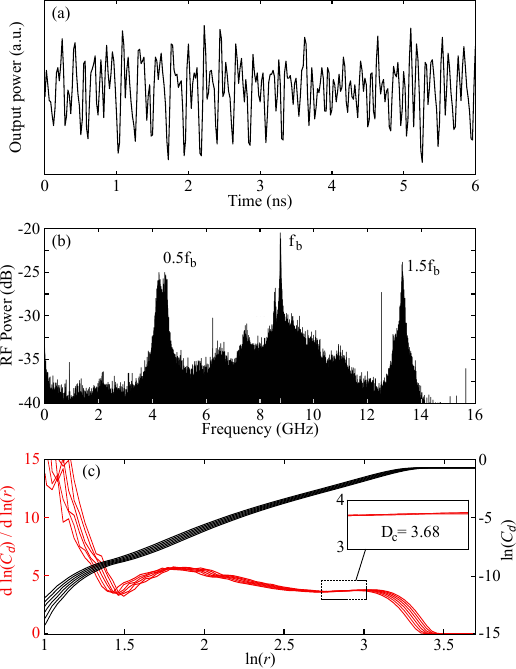}
    \caption{Chaos analysis. (a) Laser output power for $I$=4.9 mA (only the AC components of the output power are shown). (b) corresponding RF power spectrum. (c) Results of the Grassberger-Procaccia algorithm for embedding dimensions $d_e$= 25 - 30 with delay $\tau_E=0.1 ns$. The logarithmic plots of correlation integral $C_d$ versus the sphere radius $r$ is plotted in black and the slopes of the correlation integral versus $r$ in red. The convergence of the slopes with increasing embedding dimension indicates a correlation dimension $D_c$ of 3.68 (inset).}
    \label{fig_trace_temp}
\end{figure} 
\section{Chaos detection and analysis}
\begin{figure*}
    \includegraphics{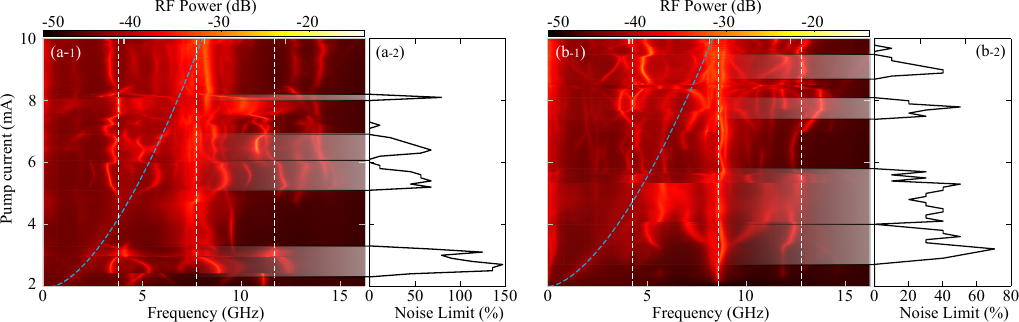}
    \caption{(a-1),(b-1) RF-spectra as function of the pump current for two other BA-VCSELs. The vertical white dashed lines indicate the birefringence frequency $f_b$ as well as 0.5$f_b$ and 1.5$f_b$. The blue curved dashed lines represent the relaxation oscillation frequencies $f_{RO}$. (a-2),(b-2) Results of the chaos titration. The shaded regions with noise limit $NL$ $>$ 0\% indicate regions of chaotic dynamics.}
    \label{fig_othervcsel}
\end{figure*}
In order to evaluate if the BA-VCSEL exhibits chaotic dynamics, we use an algorithm \cite{barahona_detection_1996} that compares a linear and nonlinear one-step prediction model of Volterra-Wiener form \cite{Volterra_nonlinear_modeling_1988} applied to the time series. This method robustly predicts deterministic chaotic dynamics, even in the presence of measurement noise or strong periodicity and with a small number of sampling point. Therefore, if the nonlinear model yields a better prediction than the linear one, it is concluded that the system exhibits chaotic dynamics. The method was later extended by incorporating a chaos titration technique, which gradually increases additive synthetic white noise \cite{poon_titration_2001, hu_chaos_titration_2006}, with the hyper-parameters associated: maximum polynomial degree $p_d=2$, maximum memory depth $\kappa=15$ and subsampling factor $M=5$.

Chaos titration provides a test of noise limit ($NL$) with decreasing signal-to-noise ratio (SNR), hence establishing the "strength" of $NL$ in the time series. The results are shown in Fig. \ref{fig_principal}(d). The "Noise Limit" is the level of added noise at which the linear model begins to outperform the nonlinear model in prediction accuracy. A positive $NL$ value indicates that the added noise is sufficient to shadow the nonlinear characteristics of the original time series, hence is indicative of chaos. The algorithm was applied to time series across the entire current range as shown in Fig. \ref{fig_principal}(d). A positive $NL$ of up to 120\% and thus chaos was found in the current regime of 3.9 mA to 5.3 mA. The edges of this region coincide with two major bifurcations, and strong frequency components near 0.5$f_b$ and 1.5$f_b$ are found inside the region. Other parameter regions with positive but much smaller $NL$ are also found for higher currents. In our laser, the number of modes increases as the pump current rises [Fig. \ref{fig_principal}(b)], leading to stronger interactions between them and an enrichment of frequency components and complexity in the RF spectrum [Fig. \ref{fig_principal}(b)]. In the application of the chaos titration method \cite{poon_titration_2001}, it has been noted that predicting chaos can be particularly challenging when the system exhibits high complexity. This can reduce the sensitivity of the detection algorithm to chaotic dynamics hence possibly explaining the smaller $NL$ achieved for higher currents.

In regions where the noise limit is zero, the linear model provides a better prediction, which can be interpreted in several ways. It may suggest that the laser does not exhibit deterministic chaos in these regions, leaving room for other possibilities such as complex mode hopping of stochastic origin. On the other hand, it is possible that deterministic chaos is present but too complex for the algorithm to detect it as chaos, perhaps due to a higher dimensionality or complexity of the laser dynamics.
Therefore, while we can be fairly confident in the presence of chaos in the detected regions, we cannot rule out the possibility of more complex chaotic dynamics at higher current levels.

We now focus on the chaotic region in the range of current between 3.9 mA and 5.3 mA to determine the dimension of the chaotic attractor using the Grassberger-Procaccia (GP) algorithm \cite{FRAEDRICH_correlationdimension_1993,procaccia_characterization_1983}. The result for $I$=4.9 mA is shown in Fig. \ref{fig_trace_temp}. The chaotic laser output power and its corresponding power spectrum are plotted in Fig. \ref{fig_trace_temp}(a),(b) respectively. Figure \ref{fig_trace_temp}(c) shows the dependency of the correlation integral $C_d$ versus the sphere radius $r$ and with embedding dimension $d_e$ ranging from 25 to 30 and an embedding delay of $t_E = 0.1ns$. The slopes of the correlation integrals (in red) converge to $D_c \simeq$ 3.68 indicated by the inset, which is the fractal dimension of the chaotic attractor at 4.9 mA. The convergence of the correlation dimension when increasing the embedding dimension is another confirmation for deterministic chaos, and does not occur for a dynamics drive by noise.

The same algorithm was used to estimate the correlation dimension of chaos from a single-mode VCSEL, and correlation dimensions up to $D_c \simeq $1.82 have been reported \cite{virte_deterministic_2013} . This emphasizes the increased complexity of our system in comparison to single-mode VCSELs, for which the lower correlation dimension indicates a relatively simple chaotic dynamics. The multimode nature of our system introduces additional degrees of freedom and the complex interplay between multiple transverse modes and polarization dynamics manifests in a higher correlation dimension of the chaotic attractor of 3.68. Comparably high complexity can be advantageous for chaos-based applications such as chaos cryptography.

A total of five BA-VCSELs of the same model were analyzed in detail. The RF-spectra and noise titration results for two of these lasers are shown in Fig. \ref{fig_othervcsel}. In each of the RF-spectra, the birefringence frequency $f_b$ and its subharmonics, as well as the relaxation oscillation frequencies are indicated, which slightly vary from one VCSEL to the other. As observed earlier, most of the frequency components are close to these characteristic frequencies, with other additional frequency components in some pump current ranges.

When applying the chaos titration analysis, we observe multiple chaotic zones at different currents, each with varying widths and noise limits, reaching up to 140\% for the most prominent region. Although all bifurcations lead to changes in the RF spectrum and in the excited modes, there does not seem to be a simple correlation between the number of modes and the complexity of the dynamics. 


The studies were further expanded by investigating the impact of temperature on our systems. In addition to shifting the wavelengths of all modes as a result of the thermal red-shift, the regions where the frequency distribution changes in the RF spectrum and the polarization switching points (PSPs) are also shifted, but with no changes in the main polarization axes $u$ and $v$. These findings suggest that the chaotic zones detected by the chaos titration are also shifted, allowing the selection of a specific chaotic region at different currents. Taking into account all the different chaotic zones, each with its unique characteristics, along with the ability to modify the number of chaotic regions and their parameter ranges, our study suggests that BA-VCSEL can be a versatile source of chaotic dynamics with superior performance compared to chaos generated by single-mode VCSELs.

\section{Conclusion}
In summary, our study of commercially available BA-VCSELs with continuous-wave pumping has revealed complex behavior, highlighting the importance of both polarization and transverse mode competition for the dynamics. The RF-spectra exhibit an increasing number of frequency components, suggesting hihly complex and possibly chaotic dynamics. We apply two methods of time series analysis, chaos titration and correlation dimension analysis in order to characterize the dynamics. These analyses reveal that some bifurcations of the dynamics lead to deterministic chaos. Caution should be applied regarding potential  false negatives of the chaos titration analysis, as certain pump current regions might exhibit high-dimensional chaos that the chaos titration algorithm could not properly detect.

We investigated several VCSELs of the same model, and all of them have at least one zone with chaotic dynamics confirmed by the chaos titration, with some VCSELs featuring several zones. In addition, we determined the correlation dimension of the reconstructed chaotic attractor, which reaches values in the range of 3.2 to 4.2. This not only confirms the chaotic nature of the dynamics, but also demonstrates that the chaotic dynamics of BA-VCSELs can be more complex that that of single-mode VCSELs for which correlation dimensions of 1.8 were found. These experimental results for a set of five BA-VCSELs demonstrate the rich and diverse dynamic behavior of this type of laser, capable of exhibiting chaotic dynamics that can be tuned through laser parameters such as temperature and current. We believe our results have significant implications for applications requiring compact sources of ultrafast chaotic dynamics such as secure communications. Since these multimode VCSELs exhibit chaotic dynamics without any need for delayed feedback, optical injections or modulation, the could be used as compact and reliable chaos generators. Futhemore, the ability to change these dynamics with simple operation parameters like pump current and temperature opens up new possibilities for the practical use of BA-VCSELs in chaos-based application.

\newpage
\textbf{Funding} The Chair in Photonics is supported by Region Grand Est, GDI Simulation, Departement de la Moselle, European Regional Development Fund, CentraleSupelec, Fondation CentraleSupelec, and Eurometropole de Metz. \\ 
\textbf{Data Availability Statement} Data underlying the results presented in this paper are
not publicly available at this time but may be obtained from the authors upon
reasonable request. \\ 
\textbf{Disclosures} The authors declare no conflicts of interest.

\bibliography{apssamp}
\end{document}